\documentclass[aps,prd,twocolumn,showpacs,superscriptaddress,groupedaddress]{revtex4}  
\usepackage{graphicx}  
\usepackage{dcolumn}   
\usepackage{bm}        
\usepackage{amssymb}   
\usepackage{amsmath}
\usepackage{bm}

\newcommand{\aOw}{\ensuremath{a_0^W}}
\newcommand{\aOz}{\ensuremath{a_0^Z}}
\newcommand{\aCw}{\ensuremath{a_C^W}}
\newcommand{\aCz}{\ensuremath{a_C^Z}}
\newcommand{\aOwL}{\ensuremath{a_0^W/\Lambda^2}}

\newcommand{\aCwL}{\ensuremath{a_C^W/\Lambda^2}}

\newcommand{\Lcutoff}{\ensuremath{\Lambda_{\text{cutoff}}}}

\newcommand{\Eslash}{\mbox{$\rm E \kern-0.6em\slash$}}
\def \etmiss {\mbox{\ensuremath{E\kern-0.6em\slash_T}}}

\newcommand{\GeV} {\ensuremath{\mathrm{Ge\kern -0.1em V}}}
\newcommand{\TeV} {\ensuremath{\mathrm{Te\kern -0.1em V}}}

\newcommand{\ttbar}{\mbox{$t\overline{t}$}}

\newcommand{\pt}{\ensuremath{p_T}}

\newcommand{\DO}{D0}
\newcommand{\WWgg}{\ensuremath{WW\gamma\gamma}}

\graphicspath{{figs/}}

\def\mytwocolumn{1}

\def \nL0  {N$_{\mathrm{L0}}$}
\begin{document}

\hspace{5.2in} \mbox{FERMILAB-PUB-13-133-E}

\title{\boldmath Search for anomalous quartic \WWgg{} couplings in dielectron and missing energy 
final states in
$\boldsymbol{p\bar{p}}$ collisions at $\boldsymbol{\sqrt{s} =}$
1.96~TeV}
\affiliation{LAFEX, Centro Brasileiro de Pesquisas F\'{i}sicas, Rio de Janeiro, Brazil}
\affiliation{Universidade do Estado do Rio de Janeiro, Rio de Janeiro, Brazil}
\affiliation{Universidade Federal do ABC, Santo Andr\'e, Brazil}
\affiliation{University of Science and Technology of China, Hefei, People's Republic of China}
\affiliation{Universidad de los Andes, Bogot\'a, Colombia}
\affiliation{Charles University, Faculty of Mathematics and Physics, Center for Particle Physics, Prague, Czech Republic}
\affiliation{Czech Technical University in Prague, Prague, Czech Republic}
\affiliation{Institute of Physics, Academy of Sciences of the Czech Republic, Prague, Czech Republic}
\affiliation{Universidad San Francisco de Quito, Quito, Ecuador}
\affiliation{LPC, Universit\'e Blaise Pascal, CNRS/IN2P3, Clermont, France}
\affiliation{LPSC, Universit\'e Joseph Fourier Grenoble 1, CNRS/IN2P3, Institut National Polytechnique de Grenoble, Grenoble, France}
\affiliation{CPPM, Aix-Marseille Universit\'e, CNRS/IN2P3, Marseille, France}
\affiliation{LAL, Universit\'e Paris-Sud, CNRS/IN2P3, Orsay, France}
\affiliation{LPNHE, Universit\'es Paris VI and VII, CNRS/IN2P3, Paris, France}
\affiliation{CEA, Irfu, SPP, Saclay, France}
\affiliation{IPHC, Universit\'e de Strasbourg, CNRS/IN2P3, Strasbourg, France}
\affiliation{IPNL, Universit\'e Lyon 1, CNRS/IN2P3, Villeurbanne, France and Universit\'e de Lyon, Lyon, France}
\affiliation{III. Physikalisches Institut A, RWTH Aachen University, Aachen, Germany}
\affiliation{Physikalisches Institut, Universit\"at Freiburg, Freiburg, Germany}
\affiliation{II. Physikalisches Institut, Georg-August-Universit\"at G\"ottingen, G\"ottingen, Germany}
\affiliation{Institut f\"ur Physik, Universit\"at Mainz, Mainz, Germany}
\affiliation{Ludwig-Maximilians-Universit\"at M\"unchen, M\"unchen, Germany}
\affiliation{Panjab University, Chandigarh, India}
\affiliation{Delhi University, Delhi, India}
\affiliation{Tata Institute of Fundamental Research, Mumbai, India}
\affiliation{University College Dublin, Dublin, Ireland}
\affiliation{Korea Detector Laboratory, Korea University, Seoul, Korea}
\affiliation{CINVESTAV, Mexico City, Mexico}
\affiliation{Nikhef, Science Park, Amsterdam, the Netherlands}
\affiliation{Radboud University Nijmegen, Nijmegen, the Netherlands}
\affiliation{Joint Institute for Nuclear Research, Dubna, Russia}
\affiliation{Institute for Theoretical and Experimental Physics, Moscow, Russia}
\affiliation{Moscow State University, Moscow, Russia}
\affiliation{Institute for High Energy Physics, Protvino, Russia}
\affiliation{Petersburg Nuclear Physics Institute, St. Petersburg, Russia}
\affiliation{Instituci\'{o} Catalana de Recerca i Estudis Avan\c{c}ats (ICREA) and Institut de F\'{i}sica d'Altes Energies (IFAE), Barcelona, Spain}
\affiliation{Uppsala University, Uppsala, Sweden}
\affiliation{Lancaster University, Lancaster LA1 4YB, United Kingdom}
\affiliation{Imperial College London, London SW7 2AZ, United Kingdom}
\affiliation{The University of Manchester, Manchester M13 9PL, United Kingdom}
\affiliation{University of Arizona, Tucson, Arizona 85721, USA}
\affiliation{University of California Riverside, Riverside, California 92521, USA}
\affiliation{Florida State University, Tallahassee, Florida 32306, USA}
\affiliation{Fermi National Accelerator Laboratory, Batavia, Illinois 60510, USA}
\affiliation{University of Illinois at Chicago, Chicago, Illinois 60607, USA}
\affiliation{Northern Illinois University, DeKalb, Illinois 60115, USA}
\affiliation{Northwestern University, Evanston, Illinois 60208, USA}
\affiliation{Indiana University, Bloomington, Indiana 47405, USA}
\affiliation{Purdue University Calumet, Hammond, Indiana 46323, USA}
\affiliation{University of Notre Dame, Notre Dame, Indiana 46556, USA}
\affiliation{Iowa State University, Ames, Iowa 50011, USA}
\affiliation{University of Kansas, Lawrence, Kansas 66045, USA}
\affiliation{Louisiana Tech University, Ruston, Louisiana 71272, USA}
\affiliation{Northeastern University, Boston, Massachusetts 02115, USA}
\affiliation{University of Michigan, Ann Arbor, Michigan 48109, USA}
\affiliation{Michigan State University, East Lansing, Michigan 48824, USA}
\affiliation{University of Mississippi, University, Mississippi 38677, USA}
\affiliation{University of Nebraska, Lincoln, Nebraska 68588, USA}
\affiliation{Rutgers University, Piscataway, New Jersey 08855, USA}
\affiliation{Princeton University, Princeton, New Jersey 08544, USA}
\affiliation{State University of New York, Buffalo, New York 14260, USA}
\affiliation{University of Rochester, Rochester, New York 14627, USA}
\affiliation{State University of New York, Stony Brook, New York 11794, USA}
\affiliation{Brookhaven National Laboratory, Upton, New York 11973, USA}
\affiliation{Langston University, Langston, Oklahoma 73050, USA}
\affiliation{University of Oklahoma, Norman, Oklahoma 73019, USA}
\affiliation{Oklahoma State University, Stillwater, Oklahoma 74078, USA}
\affiliation{Brown University, Providence, Rhode Island 02912, USA}
\affiliation{University of Texas, Arlington, Texas 76019, USA}
\affiliation{Southern Methodist University, Dallas, Texas 75275, USA}
\affiliation{Rice University, Houston, Texas 77005, USA}
\affiliation{University of Virginia, Charlottesville, Virginia 22904, USA}
\affiliation{University of Washington, Seattle, Washington 98195, USA}
\author{V.M.~Abazov} \affiliation{Joint Institute for Nuclear Research, Dubna, Russia}
\author{B.~Abbott} \affiliation{University of Oklahoma, Norman, Oklahoma 73019, USA}
\author{B.S.~Acharya} \affiliation{Tata Institute of Fundamental Research, Mumbai, India}
\author{M.~Adams} \affiliation{University of Illinois at Chicago, Chicago, Illinois 60607, USA}
\author{T.~Adams} \affiliation{Florida State University, Tallahassee, Florida 32306, USA}
\author{J.P.~Agnew} \affiliation{The University of Manchester, Manchester M13 9PL, United Kingdom}
\author{G.D.~Alexeev} \affiliation{Joint Institute for Nuclear Research, Dubna, Russia}
\author{G.~Alkhazov} \affiliation{Petersburg Nuclear Physics Institute, St. Petersburg, Russia}
\author{A.~Alton$^{a}$} \affiliation{University of Michigan, Ann Arbor, Michigan 48109, USA}
\author{A.~Askew} \affiliation{Florida State University, Tallahassee, Florida 32306, USA}
% ...
%
%
\collaboration{The D0 Collaboration\footnote{with visitors from
$^{a}$Augustana College, Sioux Falls, SD, USA,
$^{b}$The University of Liverpool, Liverpool, UK,
$^{c}$DESY, Hamburg, Germany,
$^{d}$Universidad Michoacana de San Nicolas de Hidalgo, Morelia, Mexico
$^{e}$SLAC, Menlo Park, CA, USA,
$^{f}$University College London, London, UK,
$^{g}$Centro de Investigacion en Computacion - IPN, Mexico City, Mexico,
$^{h}$Universidade Estadual Paulista, S\~ao Paulo, Brazil,
$^{i}$Karlsruher Institut f\"ur Technologie (KIT) - Steinbuch Centre for Computing (SCC)
and
$^{j}$Office of Science, U.S. Department of Energy, Washington, D.C. 20585, USA.
}} \noaffiliation
\vskip 0.25cm

\date{May 6, 2013}

\begin{abstract}
We present a search for anomalous components of the quartic gauge boson coupling \WWgg{} in events with an electron, a positron and
missing transverse energy. The analyzed data correspond to 9.7\,fb$^{-1}$ of integrated luminosity collected by the D0 detector in $p\bar{p}$
collisions at $\sqrt{s}=1.96$\,TeV.
The presence of anomalous quartic gauge couplings would manifest itself as an excess of boosted $WW$ events. No such excess is found in the data,
and we set the most stringent limits to date on the anomalous coupling parameters $\aOw$ and $\aCw$. When a form factor with $\Lcutoff=0.5$\,TeV is
used, the observed upper limits at 95\% C.L. are $|\aOwL|<0.0025$~GeV$^{-2}$ and $|\aCwL|<0.0092$~GeV$^{-2}$.
\end{abstract}

\pacs{14.70.Fm,12.60.Cn,13.85.Qk}
\maketitle

\section{Introduction}\label{sec:introduction}

In the standard model (SM) of particle physics, the couplings of fermions and 
gauge bosons are constrained by the gauge symmetries of the Lagrangian. 
The non-abelian gauge nature of the SM predicts the existence of trilinear ($VVV$) and quartic ($VVVV$) gauge couplings ($V=\gamma,W,Z$). These
include quartic couplings $\WWgg$ between $W$ bosons and photons that can be probed directly at hadron colliders~\cite{us1,us2,us3}, but that are too
small to be observed at the Tevatron, as will be shown later. Quartic couplings provide a window on
electroweak symmetry breaking~\cite{stirling1,stirling2} and can be probed by the measurement of $W$ boson pair production via two photon exchange.

Quartic couplings also allow for probing new physics that couples to 
electroweak bosons. As an example, the contribution of virtual heavy particles 
beyond the SM might manifest itself as a modification of the quartic couplings 
between $W$ bosons and photons~\cite{bsm1,bsm2,bsm3}. Observing the resulting anomalous couplings from such processes
could be the first evidence of new physics in the electroweak 
sector of the SM.

\begin{figure}
\begin{center}
\includegraphics[width=0.23\textwidth]{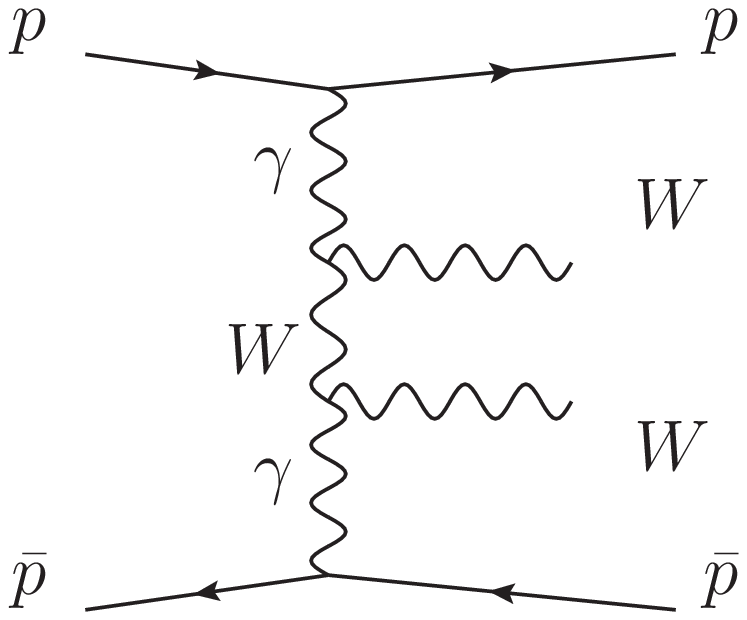}	
\hfill
\includegraphics[width=0.23\textwidth]{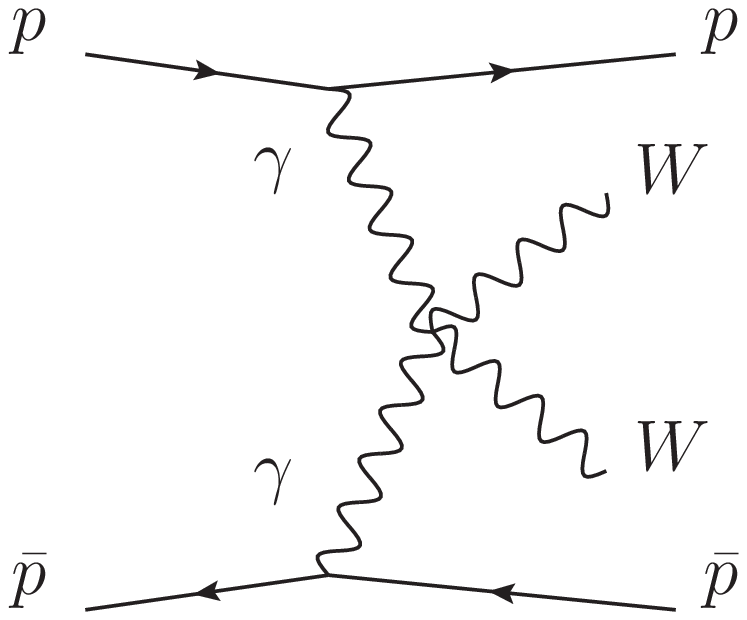}	

\unitlength=1mm
\begin{picture}(00,00)

\put(-27,-2){\text{\bf (a)}} 
\put(19,-2){\text{\bf (b)}}

\end{picture}
\end{center}
\caption{Diagrams contributing to $W$ boson pair production via photon exchange, with (a) triple $WW\gamma$ and (b) quartic \WWgg{}
couplings.}.
\label{Fig1}
\end{figure}

In this paper, we will focus on the search for \WWgg{} 
anomalous quartic gauge couplings (AQGCs) using data collected by the D0 experiment at the Fermilab $p\bar{p}$ Tevatron Collider, in events with an
electron, a positron and missing transverse energy. The main production diagrams are shown in Fig.~\ref{Fig1}. Pairs of $W$ bosons are produced via
photon exchange, where the photons are directly radiated from the colliding proton and antiproton. Triple gauge couplings $WW\gamma$ are assumed to be
at their SM values (deviations from these values have been constrained by the D0 Collaboration~\cite{d0_tgc}
and others~\cite{cdf_tgc,lep_tgc,atlas_tgc,cms_tgc}).

The parameterization of the AQGCs is
based on Ref.~\cite{Belanger:1992qh}, and only the lowest dimension operators that have
the correct Lorentz invariant structure and fulfill $SU(2)_C$ 
custodial symmetry~\cite{custodial} are considered. Such operators involving two $W$ bosons and two photons 
are of dimension six:
 \begin{eqnarray}
     \mathcal{L}_6^0 &=& \frac{-e^2}{8} \frac{\aOw}{\Lambda^2} F_{\mu\nu} 
     F^{\mu\nu} W^{+\alpha} W^-_\alpha 
     \nonumber \\
     \mathcal{L}_6^C & = & \frac{-e^2}{16} \frac{\aCw}{\Lambda^2} 
     F_{\mu\alpha} F^{\mu\beta} (W^{+\alpha} W^-_\beta + W^{-\alpha} 
     W^+_\beta),
	\label{eq:anom:lagrqgc}
\end{eqnarray}
where $F^{\mu\nu}$ is the electromagnetic field strength tensor and $W^\pm_\alpha$ is the $W^\pm$ boson field. $\aOw$ and $\aCw$ are
the usual notation for the parametrized quartic coupling constants, where a non-zero $\aOw$ could be due to an exchange of a heavy neutral scalar,
while heavy charged fermions would contribute to both $\aOw$ and $\aCw$.
The new scale $\Lambda$ is introduced so that the Lagrangian density has the correct dimension of four and is 
interpreted as the typical mass scale of  new physics. The current best 95\% C.L. limits on these anomalous parameters come from the
OPAL Collaboration from measurement of $WW\gamma$, $q\bar{q}\gamma\gamma$ and $\nu \bar{\nu}\gamma\gamma$ production~\cite{LEPlimitsQGC} at the CERN
LEP Collider:

\begin{eqnarray}
-0.020\, \GeV^{-2} < &\aOwL& < 0.020\,\GeV^{-2} \nonumber \\
-0.052\, \GeV^{-2} < &\aCwL& <0.037\,\GeV^{-2}.\nonumber\\
\label{eq:anom:qgclimits}
\end{eqnarray}

The $p\bar{p} \to p\bar{p}\,W^+W^-$ cross section via photon exchange rises quickly at 
high energies when 
the anomalous coupling parameters are non-zero, and manifests itself in particular with the production of boosted $W$ boson pairs. In the SM, the
$\gamma\gamma\to WW$ cross section is constant in the high-energy limit due to the cancellation between the relevant diagrams. When the new quartic
terms are added, the cancellation does not hold and the cross section will grow to violate unitarity at high energies. This increase of the cross
section can be
regularized with a form factor that reduces  the values of $\aOw$ and $\aCw$ at high energy while not modifying them at lower 
energies. Following a standard approach, we introduce the following form
factor~\cite{bsm1}:
\begin{eqnarray}
a^W_i\rightarrow \frac{a^W_i}{(1+M^2_{\gamma\gamma}/\Lcutoff^2)^2},
\label{eq:anom:formfactor}
\end{eqnarray}
where $M_{\gamma\gamma}$ is the invariant mass of the two photons, and $\Lcutoff$ is chosen to be either 0.5 or 1 TeV, following the
prescription of, e.g., Ref.~\cite{bsm1}. In the following, we provide limits on anomalous couplings with and without form factors.

\section{Data and Monte Carlo samples}\label{sec:datamc}

The full Run~II set of data recorded by the \DO{} detector is considered in this analysis, representing $9.7\,\text{fb}^{-1}$ of
$p\bar{p}$ collisions at $\sqrt{s}=1.96\,\text{TeV}$ delivered by the Tevatron between 2002 and 2011, after the relevant data quality requirements
are invoked. The \DO{} detector used for Run~II is described
in detail in Ref.~\cite{d0det}. 
The innermost part of the detector is composed of
a central tracking system with a silicon microstrip tracker (SMT) and
a central fiber tracker embedded within a 2~T solenoidal
magnet.  The tracking system is surrounded by a central preshower
detector and a liquid-argon/uranium calorimeter with
electromagnetic, fine, and coarse hadronic sections. The 
central calorimeter (CC) covers pseudorapidity~\cite{footnote:eta} $|\eta_d|$ $\lesssim 1.1$.
Two end calorimeters (EC) extend the coverage to $1.4\lesssim |\eta_d| \lesssim 4.2$.
Energy sampling in the region between the ECs and CC is improved by the addition of scintillating tiles.
A muon
spectrometer, with pseudorapidity coverage of $|\eta_d|\lesssim 2$, resides outside the calorimetry and is comprised of drift
tubes, scintillation counters, and toroidal magnets.
Trigger decisions are based
on information from the tracking detectors,
calorimeters, and muon spectrometer.
Details on the reconstruction and identification criteria for
electrons, jets, and missing transverse energy, \etmiss{}, can be found elsewhere~\cite{d0:hww_prd2013}. In this paper we call both electrons and
positrons ``electrons,'' with the charge of the particle determined from the curvature of the associated tracks in the central tracking system.

The background where, like the signal, the proton and the antiproton are intact in the final state, originates from photon exchange
and double pomeron exchange (DPE) processes~\cite{bib:diff}. Both these backgrounds and the AQGC signals are
modeled using the
\textsc{fpmc}~\cite{bib:fpmc} generator, followed by a detailed \textsc{geant3}-based~\cite{bib:geant} simulation of the D0 detector. Data from
random beam crossings are overlaid on the MC events to account for detector noise and additional $p\bar{p}$ interactions.
The predictions of the \textsc{fpmc} generator, which are made assuming that the proton and antiproton are left intact after the interaction, are
consistent with those of the \textsc{lpair}~\cite{bib:lpair} generator, which in turn are consistent with the measurement of the cross section for
exclusive $e^+ e^-$ production by the CDF Collaboration~\cite{bib:cdfexcl}.

Diffractive and photon exchange backgrounds to this search are exclusive $e^+ e^-$ and $\tau^+ \tau^-$ production through $t$-channel photon
exchange (Drell-Yan)
and inclusive $W^+ W^-$, $e^+ e^-$, and $\tau^+ \tau^-$ production through DPE. 

Since the outgoing intact proton and antiproton
are not detected in this measurement, we also need to consider non-diffractive backgrounds. These backgrounds are $Z/\gamma^*$+jets, $t\bar{t}$ and
diboson ($W^+W^-$, $W^\pm Z$ and $ZZ$) production, and processes in which
jets are misidentified as electrons: $W$+jets and multijet production. The simulated samples used to model them
are identical to those described in
Ref.~\cite{d0:hww_prd2013}.
All of these backgrounds, except multijet production, are
modeled using the \textsc{pythia}~\cite{bib:pythia} or \textsc{alpgen}~\cite{bib:alpgen} generator, with \textsc{pythia} providing showering and
hadronization
in the latter case, using the CTEQ6L1~\cite{bib:cteq6} parton distribution functions (PDFs). The multijet background is
determined from the data by inverting some electron selection criteria, as described in Ref.~\cite{d0:hww_prd2013}.

Single diffractive (SD) processes, for which either the incoming proton or antiproton is intact after the interaction while the other is destroyed,
have similar features to non-diffractive (ND) processes in the direction of the broken proton or antiproton, contrary to DPE processes where
both the proton and antiproton are intact. Since the cross section ratio of SD
to ND processes is about (2--3)\%, which is below the uncertainty on cross sections of ND processes cross sections, the contribution of SD processes
is neglected in
this analysis. 

The selection of data events is similar but more strict than the search for the Higgs boson in the $H \to W^+W^- \to e^+\nu e^-\bar{\nu}$ channel
that is described in detail elsewhere~\cite{d0:hww_prd2013}, which includes the same trigger approach with no explicit requirement. A
preselection is applied to the data by requiring two high-transverse momentum
(high-$p_T$) electrons with opposite charge. The leading- and trailing-$p_T$ electrons are required to satisfy $p_T^{e1}>15$\,GeV and
$p_T^{e2}>10$\,GeV, and their invariant mass is required to be $M_{ee}>15$\,GeV. In addition, these electrons are required to be within the acceptance
of the calorimeter ($|\eta_d|<1.1$ or $1.5<|\eta_d|<2.5$~\cite{footnote:eta}), with at least one electron required to be in the central part of the
calorimeter ($|\eta_d|<1.1$). The only difference from the event selection in the Higgs boson search is that we veto events with at least one
jet with $p_{T}>20$~GeV, $|\eta_d| < 2.4$, and matched to at least two tracks associated with the $p\bar{p}$ interaction vertex. The inclusive cross
section for exclusive $W$ boson pair production through photon exchange in the SM at $\sqrt{s}=2$\,TeV is $\sigma(p\bar{p}\to p\bar{p}WW) = 3$\,fb,
but after the preselection only 0.1 event is expected from this process, unless it is enhanced by AQGCs.

\begin{figure*}[!] 
\center
\includegraphics[width=0.32\textwidth]{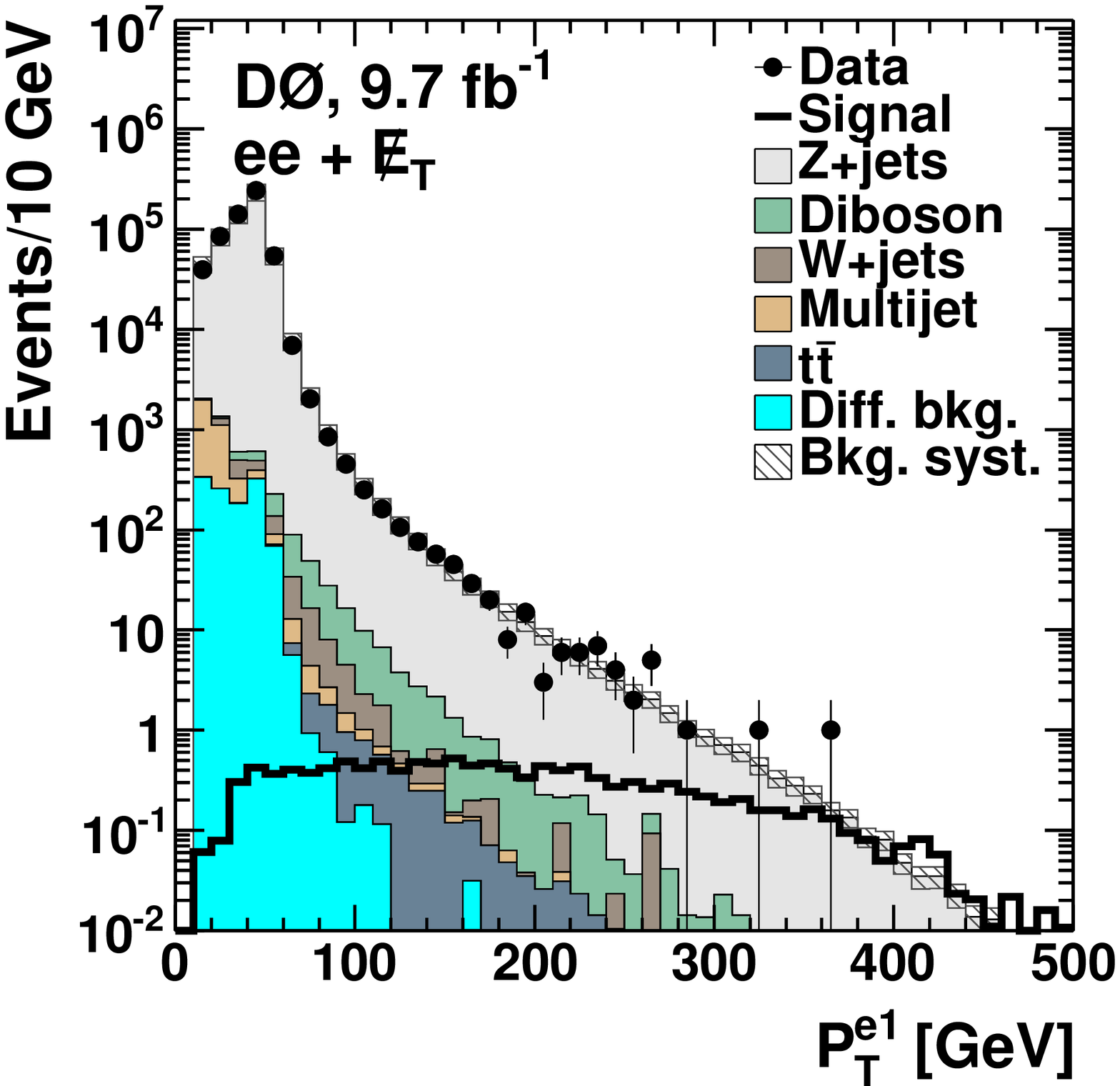}
\includegraphics[width=0.32\textwidth]{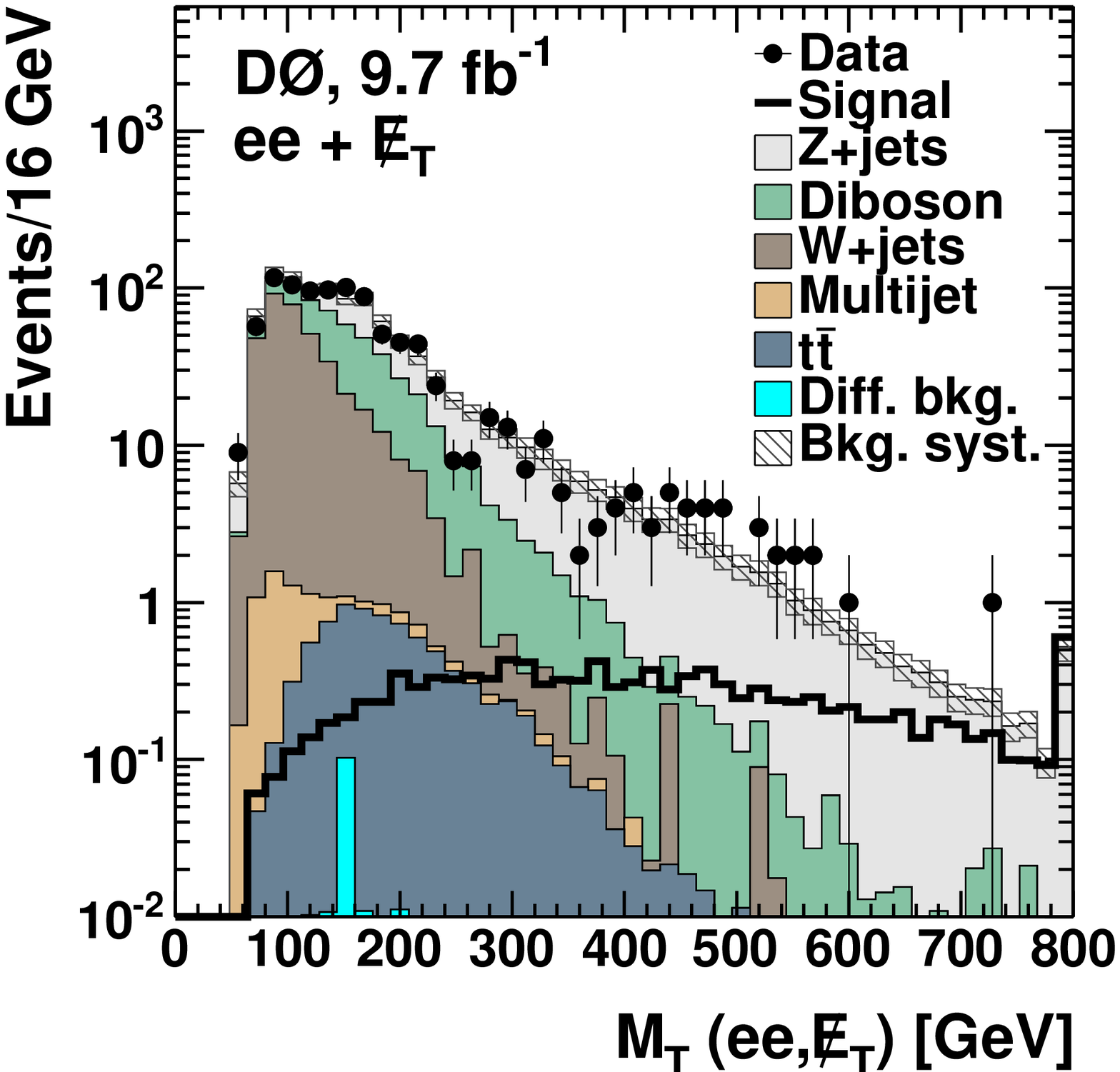}
\includegraphics[width=0.32\textwidth]{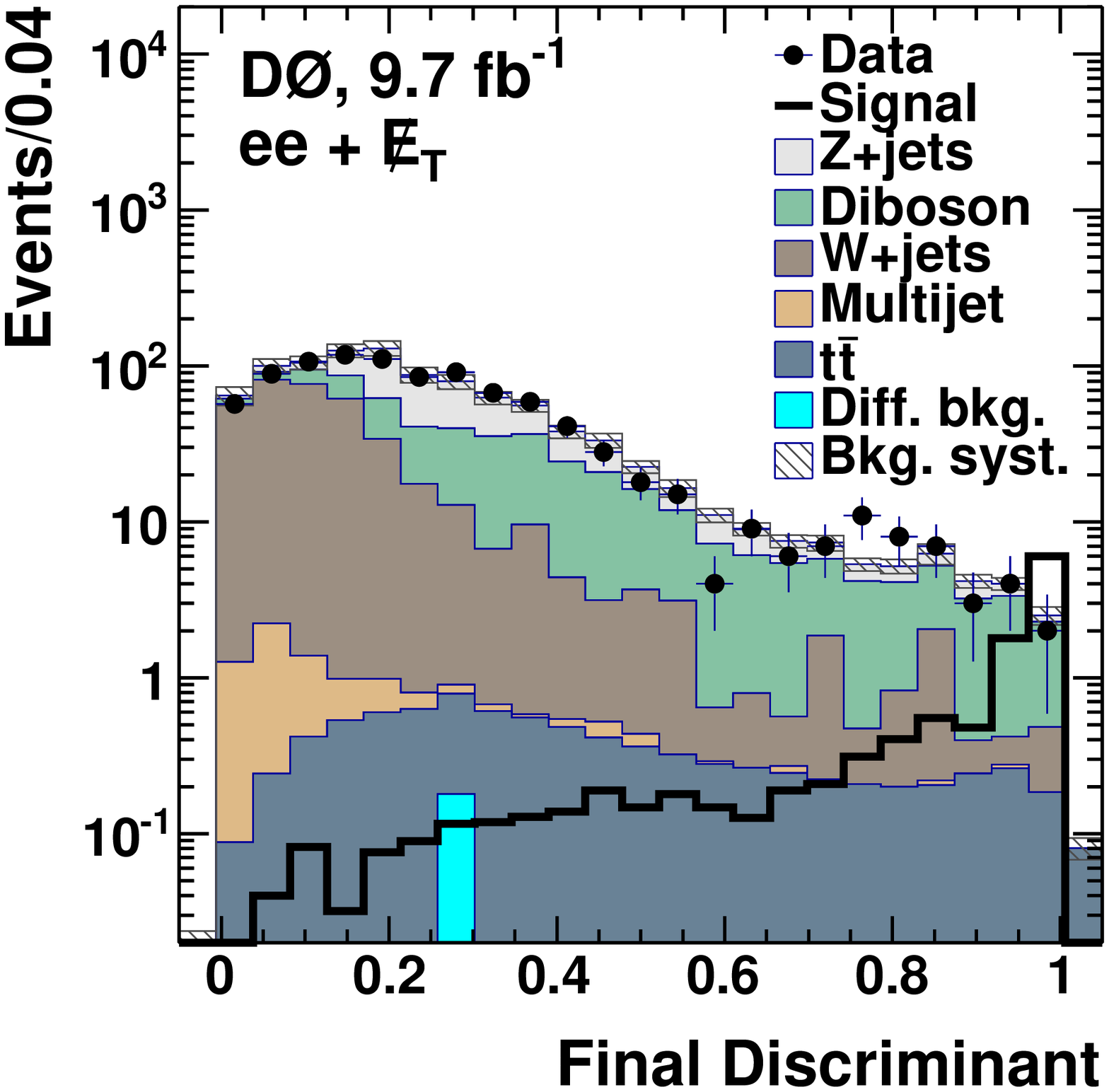}
\unitlength=1mm
\begin{picture}(00,00)

\if \mytwocolumn 1
\put(-151,45){\text{\bf (a)}} 
\put(-93,45){\text{\bf (b)}} 
\put(-34,45){\text{\bf (c)}}
\else
\put(-47,106){\text{\bf (a)}} 
\put(-17,50){\text{\bf (c)}}
\put(13,106){\text{\bf (b)}}
\fi

\end{picture}

  \caption{ [color online]
    The (a) leading electron $p_T$ at the preselection level, (b)
   the transverse mass of the \etmiss{} and the two electrons after the final selection, 
    and (c) the output of the final BDT discriminant after the final selection. In (a) and (b), the last bin includes all events
above the upper bound of the histogram.
The hatched bands show the total  systematic uncertainty on the background prediction, and
the signal distributions are those expected for $\aOwL = 5 \times 10^{-4}\,\GeV^{-2}$ and no form factor.
\label{fig:sel_all}
}

\end{figure*}

To correct for any possible mismodeling of the lepton reconstruction
and trigger efficiencies, and to reduce the impact of the luminosity
uncertainty, scale factors are applied to the Monte Carlo (MC) samples at the
preselection stage to match the data.
The $Z$ boson mass peak region in the data and MC samples after the preselection is used to 
determine normalization factors. Their differences from unity are found to be consistent with the luminosity uncertainty of 6.1\%~\cite{bib:lumi}.
The $p_{T}$ distribution of $Z$ bosons is weighted to match the distribution observed in
data~\cite{bib:zbosonpT},
taking into account its dependence on the number of reconstructed jets. 
The $p_{T}$ distribution of $W$ bosons is
weighted to match the  measured $Z$ boson $p_{T}$
spectrum,
corrected for the differences between the 
$W$ and $ Z$ $\pt$ spectra predicted in NNLO QCD~\cite{bib:w_z_bosonpT_ratio}.
The distribution of the $p_T$ of the leading electron after the preselection
is shown in
Fig.~\ref{fig:sel_all}(a).

Following the same strategy as described in Ref.~\cite{d0:hww_prd2013}, boosted decision trees (BDT) are used to reject the large $Z/\gamma^*$+jets
background, that is dominant after the preselection. The input variables to this ``selection BDT'' are kinematic quantities, including the electron
momenta, the azimuthal opening angle between the two electrons, \etmiss, variables that take into account both \etmiss{} and
its direction relative to each electron, and observables that differentiate between real and misreconstructed \etmiss. The cut on the selection BDT,
which defines the final selection, is chosen such that the contributions of the $Z/\gamma^*$+jets, $W$+jets, and diboson backgrounds are of
comparable magnitude. The distribution of the single most discriminating variable, the transverse mass of the \etmiss{} and the dielectron pair
($M_T(ee,\etmiss)=\sqrt{2\cdot p_T^{ee}\cdot\etmiss\cdot[1 - \cos\Delta\phi(ee,\etmiss )]}$), after the final selection is shown in
Fig.~\ref{fig:sel_all}(b). The expected and observed numbers of events after the preselection and the final selection are given
in Table~\ref{tab:presel_cutflow}.

A final BDT is trained to separate the AQGC signal from all the other backgrounds. The same BDT is used in the study of both
parameters $\aOw$ and $\aCw$, which feature identical kinematic characteristics. This BDT relies on the input variables of the selection BDT,
complemented with additional variables characterizing the
electron reconstruction quality to discriminate against the instrumental backgrounds (multijet and $W$+jets production). 
The distribution of the final BDT output is shown in Fig.~\ref{fig:sel_all}(c) and demonstrates the good agreement between the data and the background
expectation.

\begin{table*}[!]
\caption{\label{tab:presel_cutflow}
Observed and expected numbers of events after the preselection and the final selection for data, signal ($\aOwL = 5 \times
10^{-4}\,\GeV^{-2}$ and no form factor),
and the different backgrounds considered in the analysis  (``Diff'' stands for the diffractive backgrounds).
}
\begin{ruledtabular}

\begin{tabular}{ccrclcccccccc}
 &   \multicolumn{1}{c}{Data} & \multicolumn{3}{c}{Total background} & Signal &  \multicolumn{1}{c}{$Z/\gamma^\star \to ee$  }
 &  \multicolumn{1}{c} {$Z/\gamma^\star \to \tau\tau$} & \multicolumn{1}{c}{ $\ttbar$ }& 
\multicolumn{1}{c}{$W$+jets} &  \multicolumn{1}{c}{Diboson} &  \multicolumn{1}{c}{Multijet} & \multicolumn{1}{c}{Diff.}\\
\hline
Preselection: & 572700 & 576576 &$\pm$& 11532 & 12.2 & 566800 & 4726 & 15 & 623 & 517 & 2716 & 1180 \\
Final selection: & 946 & 983 &$\pm$& 108 & 11.6 & 291 & 22 & 8 & 370 & 287 & 5.4 & 0.2 \\
\end{tabular}

\end{ruledtabular}
\end{table*}

\section{Systematic uncertainties}
Systematic uncertainties are estimated for the signal and for each background process. They can affect
only the normalization or both the normalization and the
shape of the final discriminant. 

Sources of systematic uncertainty that affect only the
normalization arise from
the uncertainties on the theoretical cross sections of $Z$+jets ($6\%$), $W$+jets ($16\%$), diboson ($6\%$),
and $t\bar{t}$~($7\%$) processes; the multijet normalization ($30\%$); 
and the 
modeling of the $\etmiss$ for the $Z$+jets background ($5\%$). The diffractive backgrounds have been assigned a $100\%$
uncertainty on their cross sections due to the large uncertainties on the gluon density (for processes induced by pomeron exchange; the 
uncertainty on the gluon density inside the pomeron can reach 40\%, translating into an uncertainty of a factor up to 2 on the cross section) and on
the proton dissociation (for processes induced by photon exchange). For the latter process, a $20\%$ uncertainty has been assigned to the signal
theoretical cross section.

The sources of systematic uncertainty that also affect the
shape of the final discriminant distribution are quoted here as
average fractional uncertainty across bins of the final discriminant
distribution for all backgrounds: jet energy scale ($4\%$), jet
resolution ($0.5\%$), $\etmiss$ modeling (4\%), jet identification ($2\%$), jet association to the hard-scatter 
primary $p\bar{p}$ interaction vertex ($2\%$), 
and $W$+jets  modeling (10\%).
The systematic uncertainties due to the modeling of the $\pt(WW)$ and the $\Delta \phi$ between the
          leptons, and the $\pt$ of the vector boson from the $W$+jets and $Z$+jets production (see Ref.~\cite{d0:hww_prd2013}) are less than $1\%$
and taken
into account.

\section{Results}\label{sec:results}

The data are found to be in good agreement with the background-only prediction, and upper limits are set on the anomalous parameters $\aOw$ and
$\aCw$. The modified frequentist $CL_s$ method~\cite{bib:modified_freq} is
employed to set limits on the AQGCs,
where the test statistic is a log-likelihood ratio (LLR)
for the background-only and signal+background hypotheses.
The LLR is obtained by summing the LLR values of the bins of the final BDT output.
In the LLR calculation, the signal and background rates
are functions of the systematic uncertainties that are taken into
account as nuisance parameters with Gaussian priors.
Their degrading effect is reduced by
fitting the expected contributions to the data by 
maximizing the profile likelihood function for the background-only and
signal+background hypotheses separately, appropriately
taking into account all correlations between the systematic
uncertainties~\cite{bib:collie}.

\begin{figure*}[!] 
\center
\includegraphics[width=0.32\textwidth]{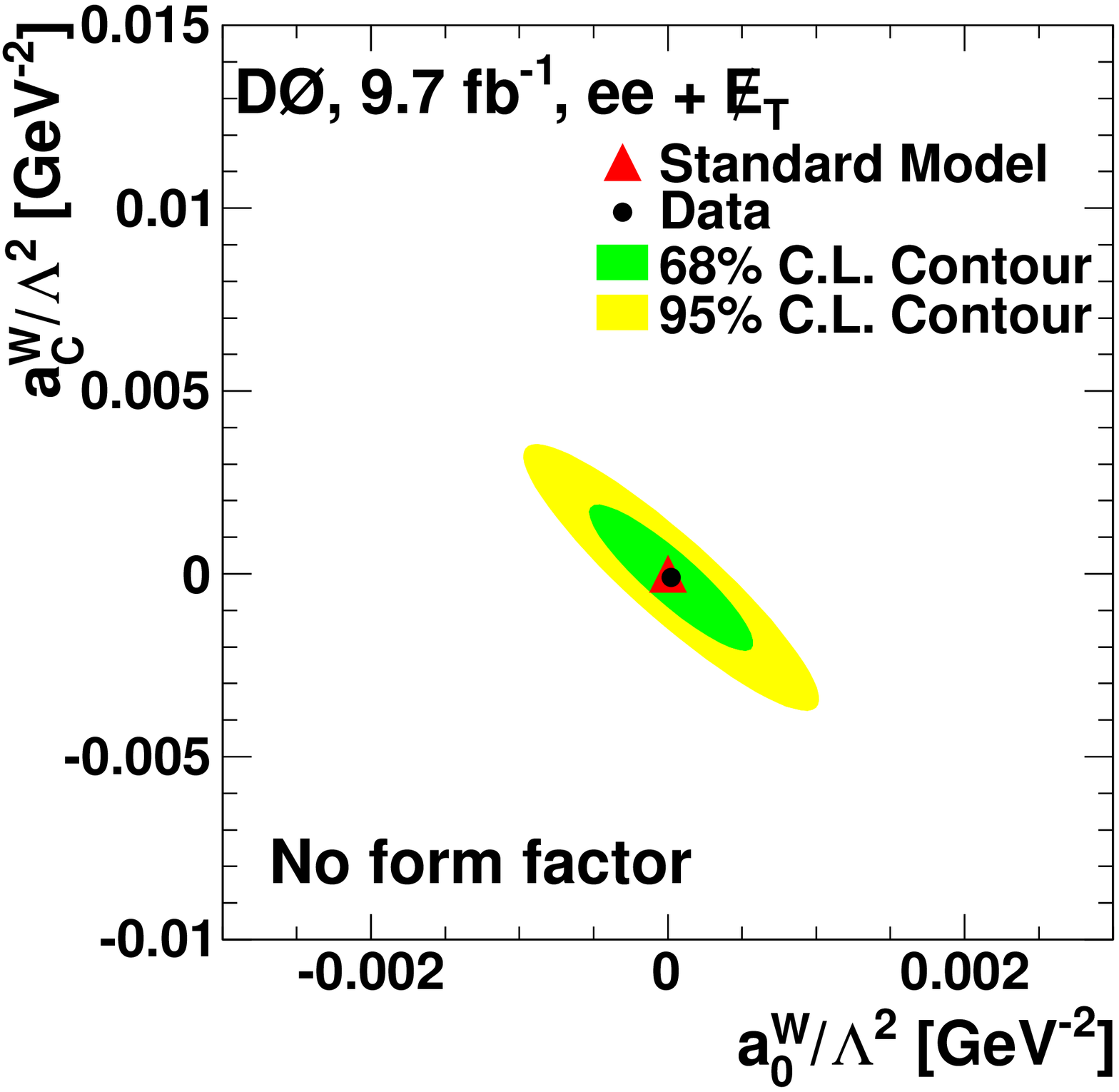}
\includegraphics[width=0.32\textwidth]{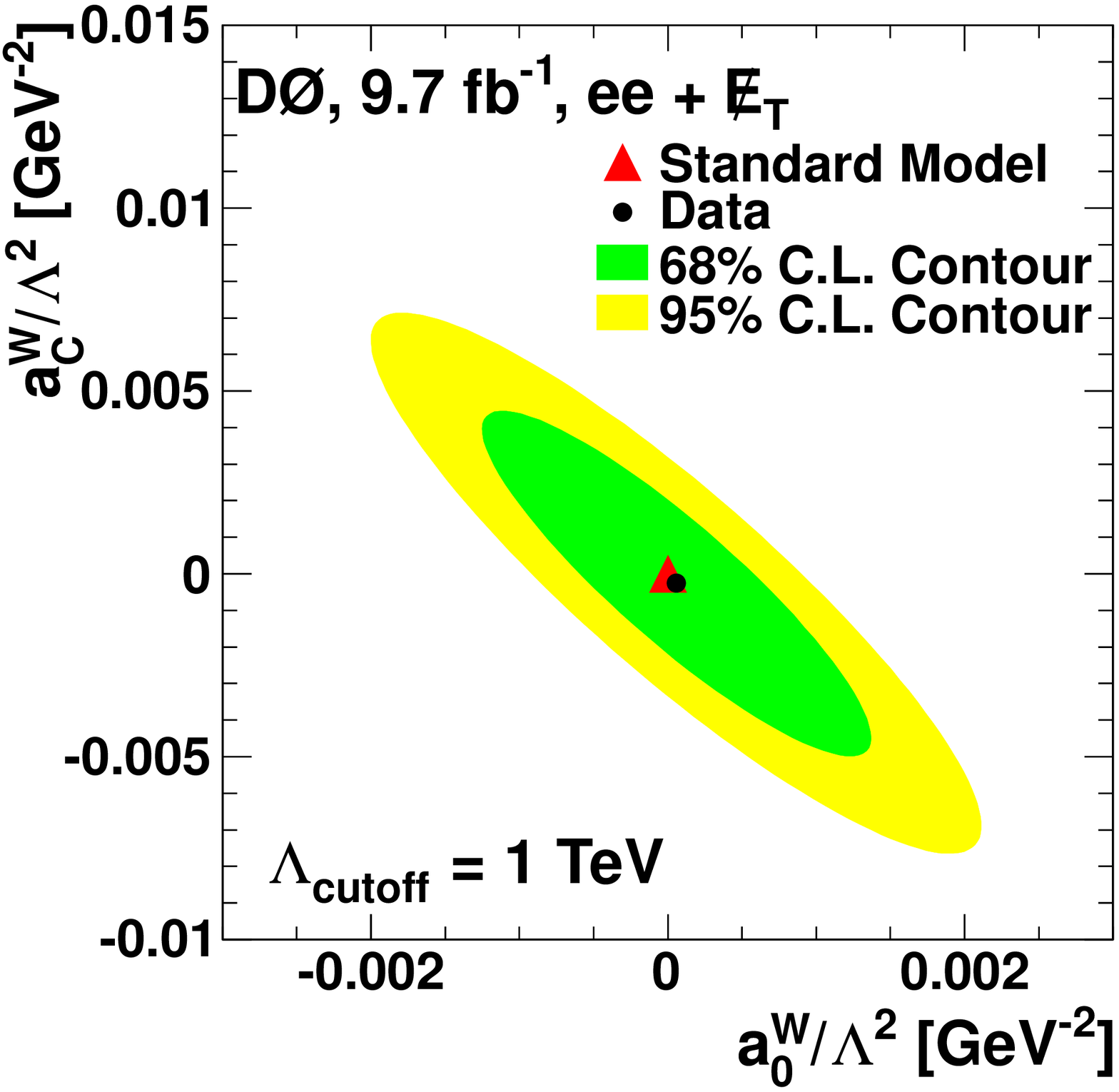}
\includegraphics[width=0.32\textwidth]{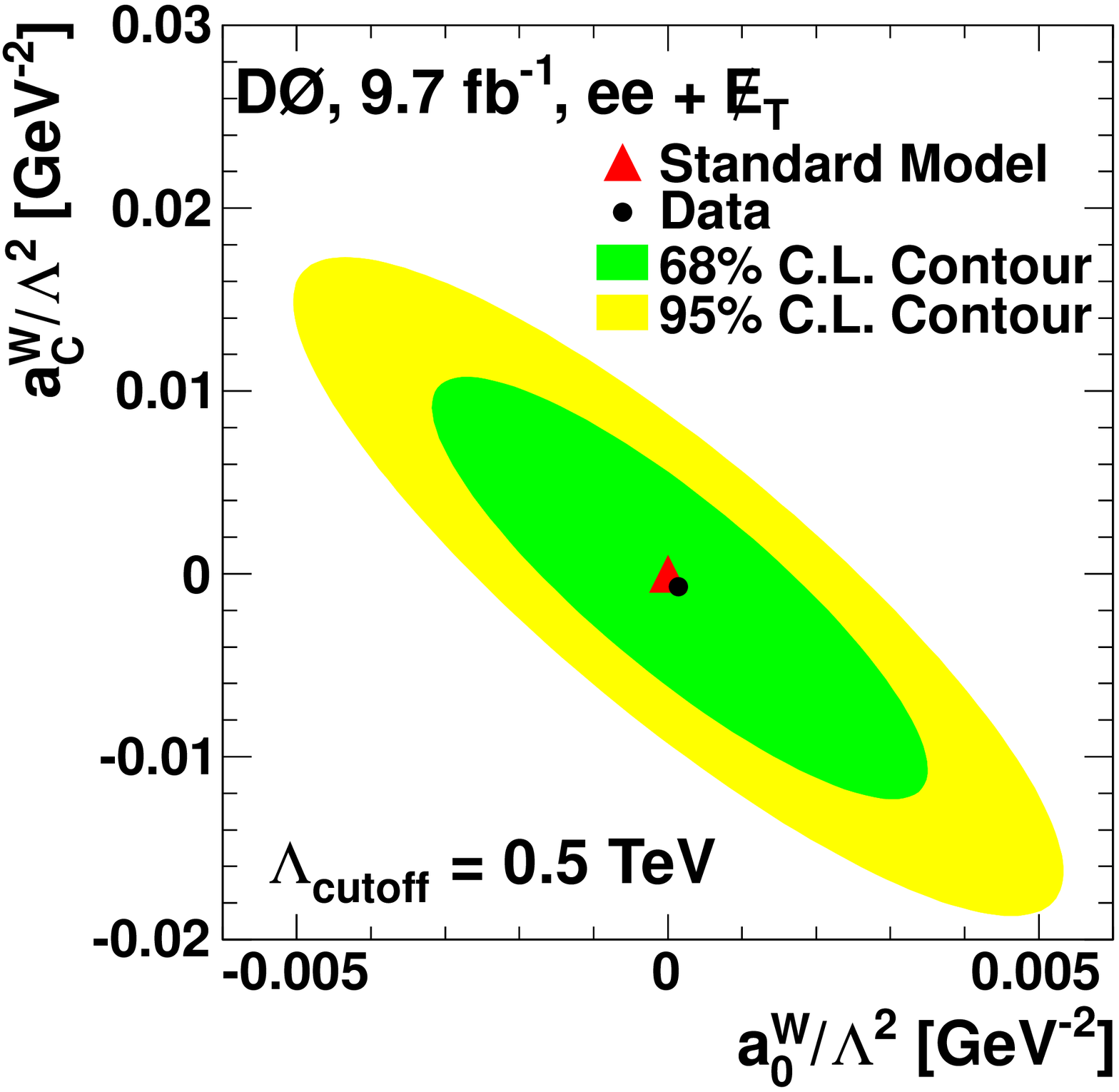}
\unitlength=1mm
\begin{picture}(00,00)

\if \mytwocolumn 1
\put(-163,44){\text{\bf (a)}} 
\put(-104,44){\text{\bf (b)}} 
\put(-45,44){\text{\bf (c)}}
\else
\put(-47,106){\text{\bf (a)}} 
\put(-17,50){\text{\bf (c)}}
\put(13,106){\text{\bf (b)}}
\fi

\end{picture}

  \caption{
  [color online] Two-parameter 68\% and 95\% C.L limits with different assumptions about the signal: (a) no form factor, or a form factor with (b)
$\Lcutoff=1$ or (c) $0.5$\,TeV.    
\label{fig:limits2d}
}

\end{figure*}

The 95\% C.L. allowed ranges for the anomalous parameter $\aOw$ ($\aCw$) can be found in
Table~\ref{tab:exclusion_a0w_fdbdt} (\ref{tab:exclusion_aCw_fdbdt}), assuming $\aCw$ ($\aOw$) is zero. The limits
are quoted both
without a form factor and for a form factor with $\Lcutoff = 1$ or $0.5$\,TeV (as advised, e.g., in Ref.~\cite{bsm1}). The
two-parameter limits are shown in Fig.~\ref{fig:limits2d} for different assumptions about the signal, namely
if no form factor is used and if a form factor is used with $\Lcutoff=1$ or $0.5$\,TeV. The two-parameter 68\%
C.L. (95\% C.L.) limits define the range of values of the
anomalous coupling parameters for which the theoretical cross section is lower than the upper 68\% C.L. (95\% C.L.) limit on the signal cross
section, obtained in the single parameter limits.
The effect of the presence of a Higgs boson with
$M_H=125$\,GeV is not accounted for, but is expected to contribute less than 4 events after the final selection, having kinematic distributions
distinct from signal, and to broaden the allowed ranges
for the anomalous parameters by a negligible amount.

\begin{table*}[htb]
\caption{Expected and observed 95\% C.L upper limits on $|\aOwL|$, assuming $\aCw$ is zero and for different assumptions about the form factor.}
\begin{center}
\begin{ruledtabular}
 \begin{tabular}{lcc}
  Cutoff & Expected upper limit [GeV$^{-2}$] & Observed upper limit [GeV$^{-2}$] \\
  \hline
  No form factor & 0.00043 & 0.00043 \\
  $\Lcutoff = 1$\,TeV &  0.00092 & 0.00089 \\
  $\Lcutoff = 0.5$\,TeV & 0.0025 & 0.0025 \\
 \end{tabular}
 \end{ruledtabular}
\end{center}
\label{tab:exclusion_a0w_fdbdt}
\end{table*}

\begin{table*}[htb]
\caption{Expected and observed 95\% C.L upper limits on $|\aCwL|$, assuming $\aOw$ is zero and for different assumptions about the form factor.}
\begin{center}
\begin{ruledtabular}
 \begin{tabular}{lcc}
  Cutoff & Expected upper limit [GeV$^{-2}$] & Observed upper limit [GeV$^{-2}$] \\
  \hline
  No form factor & 0.0016 & 0.0015 \\
  $\Lcutoff = 1$\,TeV & 0.0033 & 0.0033 \\
  $\Lcutoff = 0.5$\,TeV & 0.0090  & 0.0092 \\
 \end{tabular}
 \end{ruledtabular}
\end{center}
\label{tab:exclusion_aCw_fdbdt}
\end{table*}

\section{CONCLUSION}\label{sec:conclusion}

We have searched for anomalous \WWgg{} quartic gauge boson couplings by analyzing 9.7~fb$^{-1}$ of integrated luminosity in the $W^+W^- \to e^+\nu
e^-\bar{\nu}$ final state using the D0 detector. No excess above the background
expectation has been found. When a form factor with $\Lcutoff=0.5$\,TeV is used, the observed upper limits at 95\% C.L. are
$|\aOwL|<0.0025$~GeV$^{-2}$ and $|\aCwL|<0.0092$~GeV$^{-2}$. These are a factor 4 to 8 more stringent constraints on $\aOw$ and $\aCw$ than the
previous limits~\cite{LEPlimitsQGC}, and the only published limits to date from a hadron collider.

%
We thank the staffs at Fermilab and collaborating institutions,
and acknowledge support from the
DOE and NSF (USA);
CEA and CNRS/IN2P3 (France);
MON, NRC KI and RFBR (Russia);
CNPq, FAPERJ, FAPESP and FUNDUNESP (Brazil);
DAE and DST (India);
Colciencias (Colombia);
CONACyT (Mexico);
NRF (Korea);
FOM (The Netherlands);
STFC and the Royal Society (United Kingdom);
MSMT and GACR (Czech Republic);
BMBF and DFG (Germany);
SFI (Ireland);
The Swedish Research Council (Sweden);
and
CAS and CNSF (China).


\begin{thebibliography}{99}
\bibitem{footnote:eta}{The pseudorapidity is defined as $\eta = -\ln{(\tan\theta/2)}$, where $\theta$ is the polar angle
relative to the proton beam direction. $\eta_d$ is the detector pseudorapidity, calculated using the position of the calorimeter cluster with
respect to the center of the detector.}
\end{thebibliography}
\end{document}